\documentclass[twocolumn,superscriptaddress,aps,prllongbibliography]{revtex4-2}
\usepackage[colorlinks,bookmarks=false,citecolor=blue,linkcolor=red,urlcolor=blue]{hyperref}
\usepackage{color}
\usepackage{amsmath,amssymb,bm,braket,float,mathtools}
\usepackage{graphicx,xfrac,appendix}
\usepackage{CJK}
\usepackage{physics}
\usepackage{bbm}
\usepackage{orcidlink}
\usepackage{booktabs}


\begin{document}	
\title{Multifractality and pre-thermalization in the quasi-periodically kicked Aubry-Andr\'e-Harper model}
\author{Wen Chen\,\orcidlink{0000-0001-6089-9909}}
\email{wenchen@csrc.ac.cn}
\affiliation{Beijing Computational Science Research Center, Beijing, 100193, China}
\author{Pedro D. Sacramento\,\orcidlink{0000-0002-8276-6485}}
\email{pdss@cefema.tecnico.ulisboa.pt}
\affiliation{CeFEMA, Instituto Superior T\'{e}cnico, Universidade de Lisboa, Av. Rovisco Pais, 1049-001 Lisboa, Portugal}
\author{Rubem Mondaini\,\orcidlink{0000-0001-8005-2297}}
\email{rmondaini@csrc.ac.cn}
\affiliation{Beijing Computational Science Research Center, Beijing, 100193, China}
\begin{abstract}
    In a class of periodically driven systems, multifractal states in non-equilibrium conditions and robustness of dynamical localization when the driving is made aperiodic have received considerable attention. In this paper, we explore a family of one-dimensional Aubry-Andr\'e-Harper models that are quasi-periodically kicked with protocols following different binary quasi-periodic sequences, which can be realized in ultracold atom systems. The relationship between the systems' localization properties and the sequences' mathematical features is established utilizing the Floquet theorem and the Baker-Campbell-Hausdorff formula. We investigate the multifractality and pre-thermalization of the eigenstates of the unitary evolution operator combined with an analysis of the transport properties of initially localized wave packets. We further contend that the quasi-periodically kicked Aubry-Andr\'e-Harper model provides a rich phase diagram as the periodic case but also brings the range of parameters to observe multifractal states and pre-thermalization to a regime more amenable to experiments.
\end{abstract}
\maketitle

\section{Introduction}
The Aubry-Andr\'e-Harper (AAH) model~\cite{aubry1980analyticity,harper1955single}, whose periodicity in real space is interrupted by superimposing a lattice with incommensurate wavelengths on the primary lattice, has become an ideal platform to understand Anderson localization~\cite{anderson1958absence}. This is because, in the one-dimensional case, it is possible to observe a transition between extended-to-localized states with increased incommensurate potential strength. The system at the critical point of the transition exhibits multifractal characteristics such as neither localized nor extended self-similar wave functions, singular continuous spectra, and fractional dimension~\cite{aubry1980analyticity,harper1955single,castro2019aubry,jagannathan2021fibonacci}. Additionally, the AAH model and its extensions spark an increasing amount of interest in the theoretical investigation of exotic phenomena such as the existence and classification of mobility edges~\cite{biddle2010predicted,goncalves2023renormalization,goncalves2023hidden,zhou2023exact}, topological band structure~\cite{kraus2012topological, cestari2016fate}, non-Hermitian effects~\cite{longhi2019metal}, but also experimental realizations using photonic crystals~\cite{dal2003light, verbin2013observation, kraus2012topological, lahini2009observation}, ultracold atomic systems~\cite{Modugno2010anderson, roati2008anderson, tengxiao2021observation}, cavity-polariton devices~\cite{goblot2020emergence}, and programmable quantum superconducting processors~\cite{li2023observation}.

Departing from equilibrium, the periodically kicked AAH (kAAH) model, which depicts a tight-binding lattice periodically subjected to an instantaneous incommensurate pulse~\cite{qinpinquan2014dynamical,cadez2017dynamical}, offers a richer phase diagram, including a vast multifractal region in the space of parameters. Furthermore, it allows the exploration of dynamical localization and multifractality in non-equilibrium conditions compared to the time-independent (static) AAH model, in which the critical point must be fine-tuned. In this case, the time-independent effective Hamiltonian derived by the Floquet theorem~\cite{milena1998driven,bukov2015universal} can be used to illustrate the extended-to-localized transition in the high-frequency and small kick strength limits.

Nonetheless, if a subtle time-aperiodicity in the kicks is present~\cite{otteffect1984,cohenquantum1991}, dynamical localization disappears~\cite{cadez2017dynamical}. If white noise is added to the kick time, kick strength, or some kicks are randomly missed~\cite{ravindranath2021dynamical}, the area where dynamical localization resides is compressed, and initially localized states ultimately spread in position space. The rest of the phase diagram are a mixture of multifractal and extended states. Such instability of dynamical localization is especially relevant to experiments, where imperfections naturally abound, that require fine-tuning the time-perturbation for its observation at longer times~\cite{bitter2016experimental,sarkar2017nonexponential}.

Recently, a class of quasi-periodic sequences~\cite{barber2008aperiodic}, of which the Fibonacci and Thue-Morse are the primary representatives, has been introduced to time-dependent driven models~\cite{nandy2017aperiodically,dumitrescu2018logarithmically,ray2019dynamics,maity2019fibonacci,mukherjee2020restoring,zhangpengfei2020periodically,zhaohongzhen2021random, zhaohongzheng2022localization,tiwari2023dynamical,das2023periodically,pilatowskycameo2023complete}, such as quantum random walks~\cite{pires2020quantum}, quantum kicked rotor~\cite{bhattacharjee2022quasilocalization}, including experimental platforms of driven trapped ions~\cite{dumitrescu2022dynamical}. A common trend in those results is the presence of a phenomenon known as pre-thermalization, which describes a state that evades heating for exponentially long times~\footnote{For some quasi-periodic sequences, such as random multipolar drives, the pre-thermalization lifetimes have been argued to be algebraic~\cite{mori2021rigorous}.}, unlike randomly driven systems, which quickly result in thermalization to featureless, infinite temperature states, and has been shown to occur in various cases~\cite{dumitrescu2018logarithmically, zhaohongzhen2021random, mori2021rigorous, bhattacharjee2022quasilocalization, zhaohongzheng2022localization}. 

Following this, we examine a family of one-dimensional quasi-periodically kAAH models with kick protocols from various binary quasi-periodic sequences. We show that a link between the localization properties of the quasi-periodically kAAH model and the mathematical aspects of the sequences, including its complexity, can be established through the Floquet theorem and the Baker-Campbell-Hausdorff (BCH) formula. Additionally, the global phase diagram can be extracted via the properties of the eigenstates of the corresponding unitary evolution operator, which can take advantage of recursion relations of quasi-periodic sequences, and also from the transport properties associated with the evolution of initially localized wave packets. 

From an experimental point of view, we notice that a recent cold-atom realization of the periodic kAAH model has been reported, wherein the development of apodized Floquet engineering techniques allowed the extension of the accessible parameter range to explore multifractal states~\cite{shimasaki2022anomalous}. As a result, a further extension to the regime of quasi-periodic drivings, which we theoretically describe, can enlarge the regime where multifractality can be potentially observed.

The rest of this paper is organized as follows. In Sec.~\ref{Hamiltonian and Drive protocol}, we introduce the quasi-periodic kAAH model and its driving protocol. Section~\ref{Mathematical properties of sequences} classifies the mathematical properties of the quasi-periodic sequences, including their complexities. In Sec.~\ref{Results}, multifractality and pre-thermalization in the quasi-periodic kAAH model are explored, and lastly, Sec.~\ref{Summary} summarizes our findings.

\section{Hamiltonian and Drive protocol}
\label{Hamiltonian and Drive protocol}
A one-dimensional periodically kAAH model~\cite{qinpinquan2014dynamical, cadez2017dynamical, shimasaki2022anomalous} is  described as,
\begin{equation}
    \hat{H} = \hat{H_0} + \lambda\sum_{m} \delta(\tau - m T) \hat{H}_1,
\end{equation}
where $\hat{H}_1$ is added instantaneously with a pulsed period $T$. $m$ is an integer, $\tau$ is the time, and $\lambda$ is the kick amplitude; $\hat{H}_0$, the kinetic energy operator, is given by
\begin{equation}
    \hat{H}_0 = -J\sum_{n=1}^L \left(\hat{c}_n^{\dagger}\hat{c}_{n+1}^{\phantom{\dagger}}+\hat{c}_{n+1}^{\dagger}\hat{c}_n^{\phantom{\dagger}}\right) \ ,
\end{equation}
 where $\hat{c}_n^{\phantom{\dagger}} (\hat{c}_n^{\dagger})$ is the annihilation (creation) operator at site $n$, $J$ is the nearest-neighbor hopping amplitude, and $L$ is the system size. The potential $\hat{H}_1$ is defined as 
\begin{equation}
    \hat{H}_1 = \sum_{n=1}^L \cos(2\pi\tilde\varphi n)\hat{c}_n^{\dagger}\hat{c}_n^{\phantom{\dagger}}
\end{equation}
where $\tilde\varphi$ is an irrational number. For simplicity, we set $\tilde\varphi$ as the inverse golden ratio $(\sqrt{5}-1)/2$, and define $J = 1$ as the energy scale. Thus, the period $T$ and the kick amplitude $\lambda$ are dimensionless and can be directly mapped to the practical experimental parameters for comparison~\cite{shimasaki2022anomalous}.

According to the Floquet theorem~\cite{milena1998driven, bukov2015universal}, the time-evolution operator over one period can be derived as
\begin{equation}
    \hat{U}(T) = \hat{U}_0\hat{U}_1 = \exp\left(-i\hat{H}_0 T\right)\exp\left(-i\lambda\hat{H}_1\right)
\end{equation}
Utilizing the BCH formula, $\exp \hat{X} \exp \hat{Y} = \exp \{\hat{X} + \hat{Y} + \frac{1}{2} [\hat{X}, \hat{Y}] + \frac{1}{12}[[\hat{X}, \hat{Y}], \hat{Y}] + ...\}$, one can write down the Floquet operator as, 
\begin{equation}
    \hat{U}(T) = \exp\left[-i \left(\hat{H}_0 +\frac{\lambda}{T} \hat{H}_1\right)T+ \hat{H}_c\right],
\end{equation}
where $\hat{H}_0 + \frac{\lambda}{T} \hat{H}_1$ is an effective AAH model with renormalized potential $\lambda_{\rm AA} = \lambda /T$ in the limit of $1/T \gg 1$ (high-frequencies) and $\lambda \ll 1$ (small kick amplitudes); $\hat{H}_c$ is a higher-order contribution, i.e., which depends in larger powers of $\lambda$ and $T$. Now, suppose one has a more complicated \textit{periodic} sequence, with $n_0$ ($n_1$) number of $\hat U_0$ ($\hat U_1$)'s at each period, then the corresponding Floquet operator reads
\begin{equation}
    \hat{U}(n_0T) = \exp\left[-i \left(n_0\hat{H}_0 +n_1\frac{\lambda}{T} \hat{H}_1\right)T+ \hat{H}_c^\prime\right]\ ,
     \label{eq:period_floquet_op}
\end{equation}
where again $\hat H_c^\prime$ is a higher-order contribution that depends on the commutation between all elements in the sequence.
In our work, we consider the drive protocol following \textit{quasi-periodic} sequences $\mathcal{S}_N = \{b_1 b_2 b_3 ... b_N\}$, where $N$ is the length of the sequence and $b_n$ can only be $0$ or $1$, corresponding to 
\begin{equation}
    \hat{U}_N = \hat{U}_{b_1}\hat{U}_{b_2}\hat{U}_{b_3}...\hat{U}_{b_N}
\end{equation}
For a sufficiently long sequence, the `Floquet' operator can always be simplified as 
\begin{equation}
    \hat{U}_N = \underbrace{\hat{U}(n_0 T) \hat{U}(n_0 T) \hat{U}(n_0 T) ...\hat{U}(n_0 T)}_{N/(n_0+n_1)} + \hat{U}^\prime \label{eq7}
\end{equation}
where we shall see is valid when the ratio $N_0/N_1$ of the number of $\hat U_0$'s to the number of $\hat U_1$'s converge, and $\hat{U}^\prime$ is created by adjusting the quasi-periodic sequence to the periodic form. Now, using Eq.~\eqref{eq:period_floquet_op}, we can simplify it to
\begin{equation}
    \hat{U}_N = \exp\left\{-i \frac{N}{n_0+n_1}\left[\left(n_0\hat{H}_0 +n_1\frac{\lambda}{T} \hat{H}_1\right)T+ \hat{H}_c^{\prime\prime}\right]\right\},
    \label{eq:qp_floquet_op}
\end{equation}
where we notice that $n_0/n_1 = N_0/N_1$ is the co-prime ratio. 

As a result, setting aside higher order terms, the quasi-periodic kAAH model can be mapped to an effective AAH model as 
\begin{equation}
    \hat{H} = \hat{H}_0 + \frac{n_1\lambda}{n_0 T}\hat{H}_1,
    \label{eq:H_eff_qp}
\end{equation}
with critical point
\begin{equation}
    \frac{n_1\lambda}{n_0 T} = \frac{N_1 \lambda}{N_0 T} = 2\ ,
\label{eq:qp_crit_point}
\end{equation}
again, in the limit of $1/T \gg 1$ and $\lambda \ll 1$. As a result, the dynamical localization characteristics of the quasi-periodic kAAH model and the mathematical properties of quasi-periodic sequences are closely connected. Initially, increasing the randomness of the sequence will result in a larger $\hat{U}'$ in Eq.~(\ref{eq7}), which will further reduce the range of parameters for which the effective Hamiltonian is well established. Second, longer sequences and imbalances between $N_0$ and $N_1$ in the sequence will lead to an increase of $\hat{H}_c^\prime$ in Eq.~(\ref{eq:period_floquet_op}), having a similar impact as the above-mentioned more random sequence. In addition, the critical point of the localized-extended transition is determined by $N_1/N_0$. Consequently, it is essential to analyze the mathematical aspects of quasi-periodic sequences, particularly their elemental ratios and randomness, to comprehend the dynamical localization and multifractality of quasi-periodic kAAH models.

\section{Mathematical properties of sequences}
\label{Mathematical properties of sequences}
\begin{figure}
    \centering
    \includegraphics[width=0.99\columnwidth]{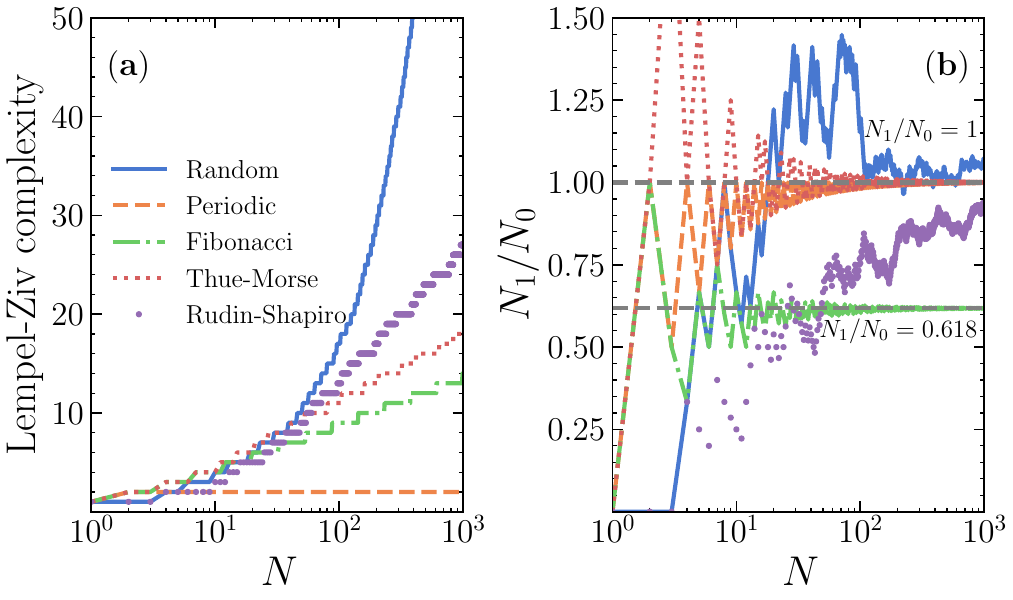}
    \caption{({\bf a}) Dependence of the Lempel-Ziv complexity with the sequence length $N$ of representative quasi-periodic sequences (Fibonacci, Thue-Morse, and Rudin-Shapiro) compared with the periodic and random cases. ({\bf b}) The same for the fraction $N_1/N_0$ of the number of elements of each type. Apart from the Fibonacci, all sequences exhibit $N_1/N_0 \to 1$ in the large $N$-limit. The Fibonacci sequence reaches $N_1/N_0 \to \tilde\varphi = \frac{\sqrt{5} - 1}{2}$ instead --- the horizontal dashed lines mark these asymptotic values.}
    \label{Fig1}
\end{figure}

We start by describing the generation processes for three representative quasi-periodic sequences: Fibonacci, Thue-Morse, and Rudin-Shapiro. By substituting elements in the initial sequence $\mathcal{S}_0 = \{b_0\}$, $b_0 = 0$, with new ones and iterating the process while following the specific rules as follows, binary quasi-periodic sequences can be produced~\cite{barber2008aperiodic}, 
\begin{itemize}
    \item Fibonacci: $0\to 01$ and $1\to 0$.
    \item Thue-Morse: $0\to01$ and $1\to 10$.
    \item Rudin-Shapiro: first, we set $A\to AB$, $B\to AC$, $C\to DB$, and $D\to DC$, then we set $A=B\to 0$ and $C=D\to 1$.
\end{itemize} 
where $N$ is the sequence length. For instance, the Thue-Morse sequence is obtained from the process: $0\to 01 \to 0110 \to 01101001 \to 0110100110010110 \to \ldots$.

To characterize the degree of complexity of the different quasi-periodic sequences of length $N$, we compute their Lempel-Ziv complexity~\cite{lempel1976on} and the fraction of its element types, $N_1/N_0$, as shown in Fig.~\ref{Fig1}. The latter is intimately tied to the dynamical localization properties, as advanced in Eqs.~\eqref{eq:H_eff_qp} and \eqref{eq:qp_crit_point}, and aids in further investigating the relation between the mathematical features of quasi-periodic sequences and the localization phenomenon.

The Lempel-Ziv complexity quantifies the amount of randomness of a given sequence by counting its number of nonidentical subgroups when it is scanned from $b_0$ to $b_N$; it can be easily computed using the Kaspar-Schuster's method~\cite{kaspar1987easily}, for example. For the case of a periodic sequence, this quantity quickly saturates with $N$ in the sequence period. For a random sequence, on the other hand, the complexity rapidly grows with its length [see Fig.~\ref{Fig1}(a)]. In turn, quasi-periodic sequences fall between these limits, but their complexity depends on the sequence type: Fibonacci, Thue-Morse, and Rudin-Shapiro describe sequences with growing complexities.

The fraction $N_1/N_0$ is also a quick and valuable metric to classify different sequences. When $N\gg1$, $N_1/N_0 \to 1$ for both periodic (with period 2) and random sequences, as shown in Fig.~\ref{Fig1}(b), i.e., the number of zeros and ones in a long periodic or random sequence is equal. The latter requires a larger length to have the same amount of zeros and ones compared to a periodic sequence due to the randomness. Similarly, $N_1/N_0$ of Rudin-Shapiro and Thue-Morse sequences approach $1$ for a sufficiently large $N$, with the latter being closer to the periodic case even at shorter sequences. Due to its lower complexity, Fibonacci quasi-periodic sequences quickly stabilize the ratio compared to the Rudin-Shapiro and Thue-Morse cases, and the value of $N_1/N_0$ converges towards the inverse golden ratio $\tilde\varphi$~\cite{ribeiro2004aperiodic}.

Many other quasi-periodic sequences can be defined, as shown in Appendix~\ref{Other sequences case}, exhibiting different degrees of complexity. In what follows, we will focus on the three examples previously listed since they are more often used in the literature and are good representatives with differing mathematical properties.

\section{Results}
\label{Results}
Having established the mathematical properties of the different sequences, we now focus on their influence on the kAAH model. In particular, we investigate the emergence of multifractality, dynamical localization in the $1/T\gg 1$, $\lambda \ll 1$ regime, and the ensuing prethermalization.

\subsection{Properties of the time-evolution operator and multifractality}
\begin{figure}
    \centering
    \includegraphics[width=0.99\columnwidth]{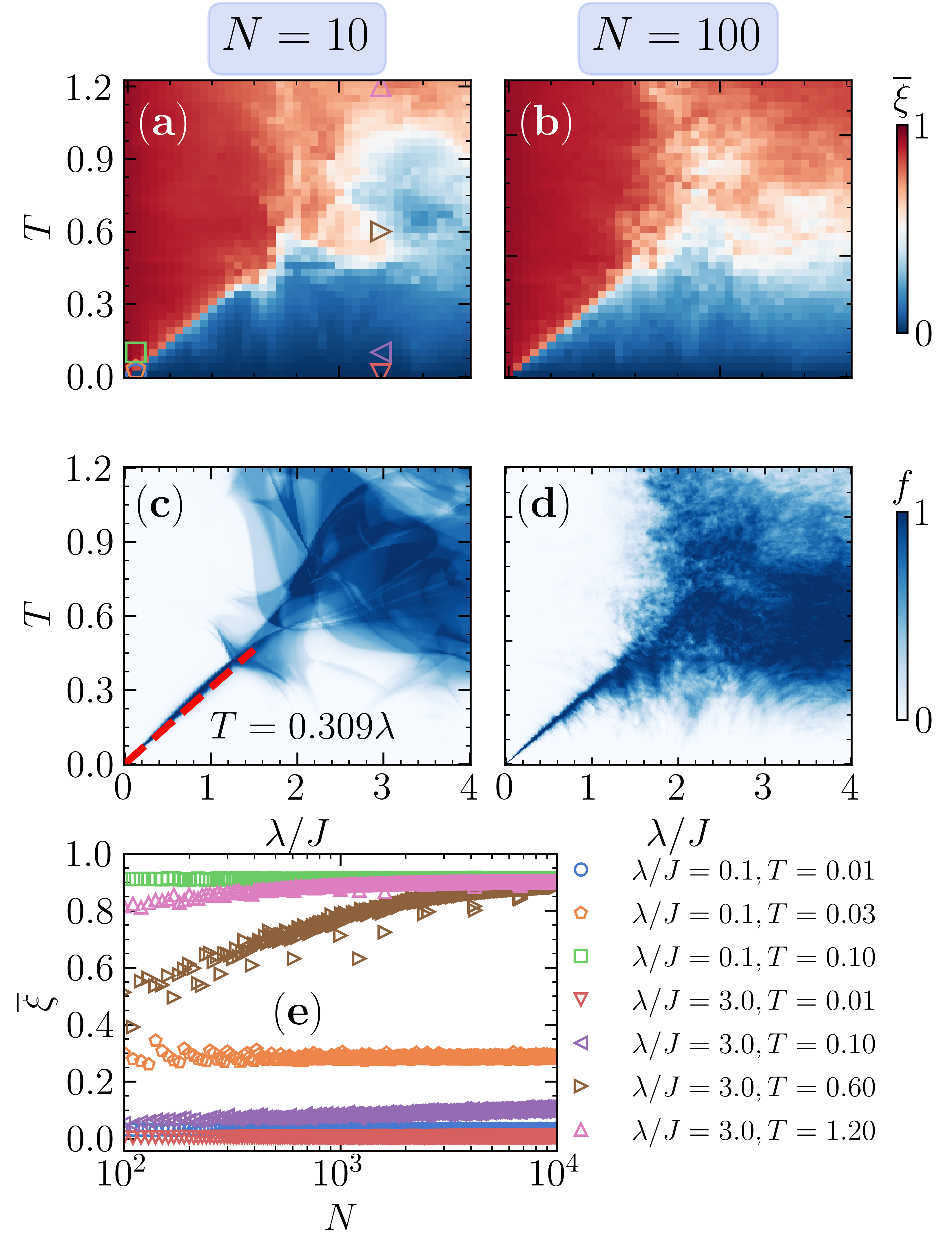}
    \caption{Phase diagram of the quasi-periodically kAAH model with Fibonacci protocol in the $\lambda$--$T$ space displaying the average IPR exponent $\overline{\xi}$ with sequence length ({\bf a}) $N=10$ and ({\bf b}) $N=100$. The corresponding fraction of multifractal states $f$ (see text) is shown in ({\bf c}) and ({\bf d}), respectively. ({\bf e}) The evolution of $\overline{\xi}$ with the length of the sequence $N$ for representative parameters as annotated in ({\bf a}), using similar color code. Here, the system size is $L = 1000$.}
    \label{Fig2}
\end{figure}

\begin{figure}
    \centering
    \includegraphics[width=0.99\columnwidth]{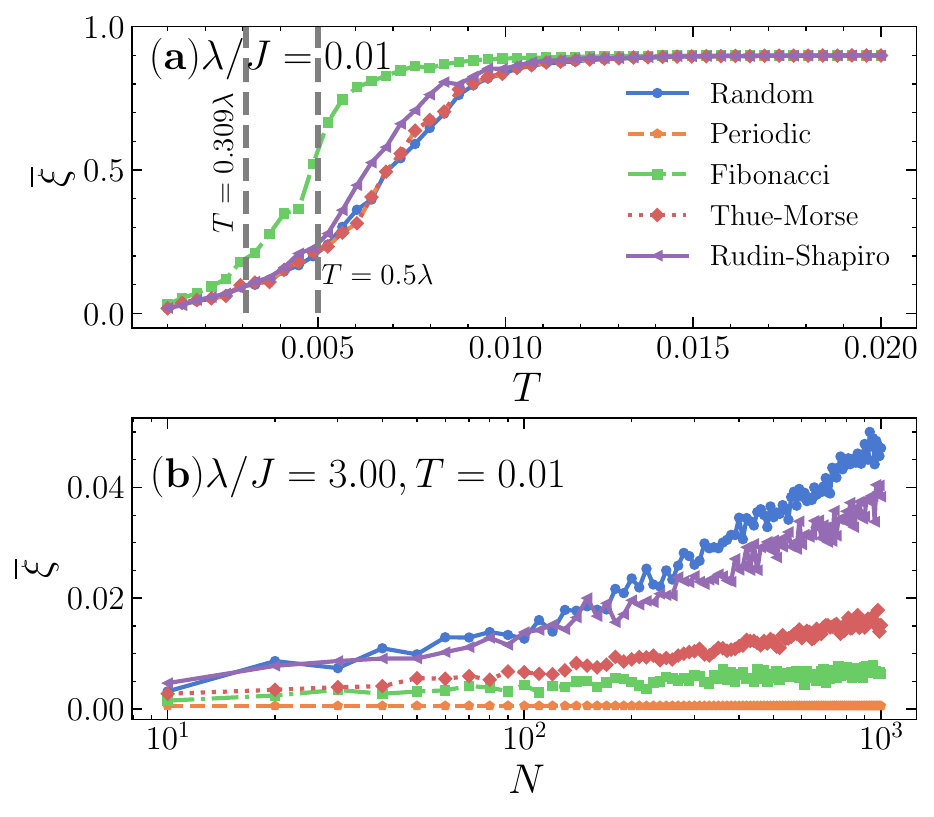}
    \caption{(\textbf{a}) Comparison of  $\overline{\xi}$ as a function of the `period' $T$ for different sequences types; here the length of the sequence $N = 1000$ for a small kick amplitude $\lambda = 0.01$. (\textbf{b}) Dependence of $\overline{\xi}$ on $N$ for the same sequences in a strong kick amplitude regime, $\lambda = 3.0$, and small period $T = 0.01$.}
    \label{Fig3}
\end{figure}

We start by classifying the complexity of the eigenstates $|\alpha\rangle$ of the unitary evolution operator $\hat U_N$ using the inverse participation ratio ${\rm IPR}^{(\alpha)} = \sum_{n=1}^{L} |\psi_n^{(\alpha)}|^4$, where $\psi_n^{(\alpha)} = \langle n|\alpha\rangle$ is the contribution of $|\alpha\rangle$ in the site-basis component $|n\rangle = \hat c_n^\dagger|\emptyset\rangle$. One can define it as ${\rm IPR}^{(\alpha)} \sim L^{-{\xi^{(\alpha)}}}$, which takes its minimum ${\rm IPR}= \frac{1}{L}\sim 0$ ($\xi \sim 1$) or maximum ${\rm IPR}=1$, ($\xi = 0$), corresponding to either extended or localized wave functions in position space, respectively. The average $\xi^{(\alpha)}$, $\overline{\xi} = (1/L)\sum_{\alpha=1}^L \xi^{(\alpha)}$, can be used to characterize the global localization properties of the system. 

Focusing on the Fibonacci quasi-periodic sequence, we report in Figs.~\ref{Fig2}(a) and \ref{Fig2}(b) the phase diagram of the time-evolution operators in the $\lambda-T$ parameter space for two sequence lengths $N =10$ and 100. We observe marked regimes with $\overline{\xi}\simeq 1$\ $({\rm or\ }0)$, denoting delocalization (or localization), but an extensive region exhibits $\overline{\xi}$ evading these limiting values. Such regions, which grow with $N$ [see Fig.~\ref{Fig2}(e) for representative points], thus suppressing dynamical localization, can be identified with the emergence of multifractality and have been the focus of recent experimental investigation in the case of periodic kicks for the current model~\cite{shimasaki2022anomalous}. Preliminary quantification of the degree of multifractality can be given by the fraction $f=N(\xi_{\rm multifractal})/L$, where $N(\xi_{\rm multifractal})$ gives the number of states with $0.2<\xi <0.8$. Figures~\ref{Fig2}(c) and \ref{Fig2}(d) more clearly display when multifractal eigenstates are likely: when both $T$ and $\lambda$ are large, but also along a line where $T \simeq 0.309 \lambda$ in the high-frequency limit.

This latter result can be quickly understood in terms of the critical point calculation developed in Sec.~\ref{Hamiltonian and Drive protocol} for the case of quasi-periodic drives:
\begin{equation}
\frac{T}{\lambda}=\frac{1}{2}\frac{N_1}{N_0}\ .
\label{eq:qp_critical_point}
\end{equation}
Since at longer sequences, $N_1/N_0$ quickly converges to $\tilde\varphi$ for a Fibonacci driving protocol, one expects thus along the line $T=(\tilde\varphi/2)\lambda \simeq 0.309\lambda$ to see critical behavior and multifractal eigenstates, precisely because in equilibrium the spectrum displays multifractality at this transition point. A similar analysis holds for other quasi-periodic sequences, noting that the different ratio $N_1/N_0$ will result in  different slopes related to the criticality at small $\lambda$ and $T$ (see Appendix~\ref{Other phase diagrams} for details).

Further characterization is given in Fig.~\ref{Fig2}(e), where we show the dependence of the average IPR exponent $\overline\xi$ on the length of the sequence $N$ for representative points in the phase diagram, as annotated in Fig.~\ref{Fig2}(a), contrasting both small ($\lambda/J=0.1$) and large ($\lambda = 3$) kick amplitudes. In the former case, $T=0.01, 0.03$, and $0.10$ are examples of localized $(\overline\xi \simeq 0.0)$, critical $(\overline\xi \simeq 0.3)$, and extended $(\overline\xi \simeq 1)$ states, respectively, with $\overline\xi$ hardly changing throughout substantially long sequences. The latter case with large kick strengths, on the other hand, shows that the behavior can be significantly altered once $N$ is large: $T=0.6$ and 1.2, which seemingly result in multifractal eigenstates at small sequence lengths, give way to extended states $(\xi \simeq 1)$ with increasing $N$. Yet, if taking short propagation times $T$ of the $\hat U_0$ operator ($T=0.01$ and 0.1) it leads to small $\overline \xi$ that very slowly grows with $N$ --- this is a first signature of prethermalization that will be investigated in Secs.~\ref{sec:prethermalization} and \ref{sec:wave_packet} in details.

Expanding this description to different sequence types, we show in Fig.~\ref{Fig3}(a) the localization-delocalization transition with growing $T$ at small kick strengths ($\lambda=0.01$) for different cases. As we previously advanced, the rapid growth of $\overline\xi$ with $T$ occurs at $T \simeq N_1\lambda/(2N_0)$. In turn, the seemingly prethermalized regime with $\overline \xi$ very slowly growing with $N$ at large kick strengths $(\lambda=3)$ is more robust once the sequence is less complex, i.e., other than the periodic drive, the quasi-periodic Fibonacci sequence displays most significant resilience to thermalization, followed by the Thue-Morse and Rudin-Shapiro sequences, directly related to the complexity analysis in Fig.~\ref{Fig1}(a).

\begin{figure}
    \centering
    \includegraphics[width=0.99\columnwidth]{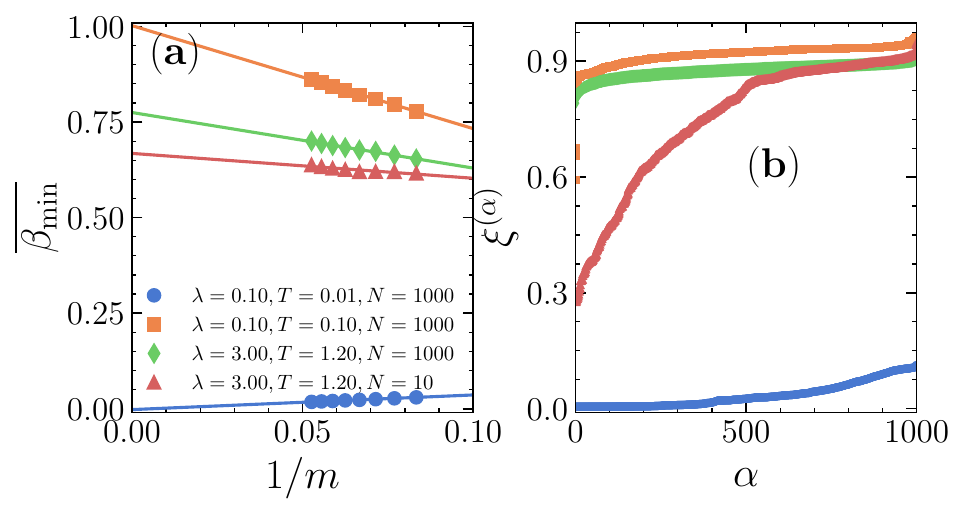}
    \caption{(\textbf{a}) $\overline{\beta_{\text{min}}}$ as a function of the inverse Fibonacci index $1/m$ at different parameters; color and markers are the same for parameters in Fig.~\ref{Fig2}(e) and lines are linear fittings. (\textbf{b}) Value of the IPR exponent $\xi^{(\alpha)}$ resolved for each eigenstate for the same parameters as in ({\bf a}). In both cases a lattice with $L=1000$ is studied.}
    \label{Fig4}
\end{figure}

An analysis of the fate of multifractality of the eigenstates of $\hat U_N$ for increasing lattice sizes $L$ can be made via a scaling process~\cite{hiramoto1989scaling, wangjun2016phase}. For instance, for a system with $L=F_m$, the $m$-th Fibonacci number, the fractional dimension $\beta_n$ can be calculated by
\begin{equation}
    p_n = F_m^{-\beta_n},
\end{equation}
where $p_n$ is the probability density at the $n$-th site. Extended, multifractal, and localized states are characterized by $\beta_{\rm{min}}$, the smallest $\beta_n$ for all sites, namely: 
(i) $\beta_{\rm{min}}=1$, extended; (ii) $0<\beta_{\rm{min}}<1$, multifractal, and (iii) $\beta_{\rm{min}}=0$, localized~\cite{hiramoto1989scaling, wangjun2016phase}. Figure~\ref{Fig4}(a) shows this scaling for the average $\beta_{\rm min}$ over the eigenstate spectrum ($\overline{\beta_{\rm min}} = \sum_\alpha \beta_{\rm min}^{(\alpha)}/F_m$). Previous analysis in Fig.~\ref{Fig2}(e) indicated that at $\lambda/J = 0.1$ in sequences with $N=10^4$, $T=0.01$ and $0.1$ led to an average IPR exponent signifying localization and delocalization, respectively, in a lattice with $L = 1000$. The scaling with the system size in Fig.~\ref{Fig4}(a) indicates that, indeed, this is confirmed in approaching the thermodynamic limit ($m\to\infty$) where $\overline{\beta_{\rm min}}\to 0$ and 1. For a case where multifractality was suggested in Fig.~\ref{Fig2}(e), $\lambda = 3$ with  $T = 1.2$, for example, we notice that the system size extrapolation of $\overline{\beta_{\rm min}}$ for $N=10$ and 1000 indeed indicates that $\lim_{m\to\infty}\overline{\beta_{\rm min}} \in (0,1)$, although one systematically approaches the extended regime for increasingly large sequence lengths.

Lastly, while we have focused on the average properties of the spectrum when describing its properties, it is important to highlight that there may be significant discrepancies when looking at individual eigenstates. We report in Fig.~\ref{Fig4}(b) the values of the IPR exponent $\xi^{(\alpha)}$ for each eigenstate for the same parameters as in Fig.~\ref{Fig4}(a) in a lattice with $L=1000$. Extended and localized cases lead to consistently large and small values of $\xi^{(\alpha)}$ across the spectrum. Still, in critical (or multifractal) regimes, the spectrum can be highly inhomogeneous with a broad range where $\xi^{(\alpha)}$ lies into. This highlights that it is often necessary to resolve spectral features when describing multifractal behavior in the quasi-periodically kAAH.

\begin{figure}[b]
    \centering
    \includegraphics[width=0.99\columnwidth]{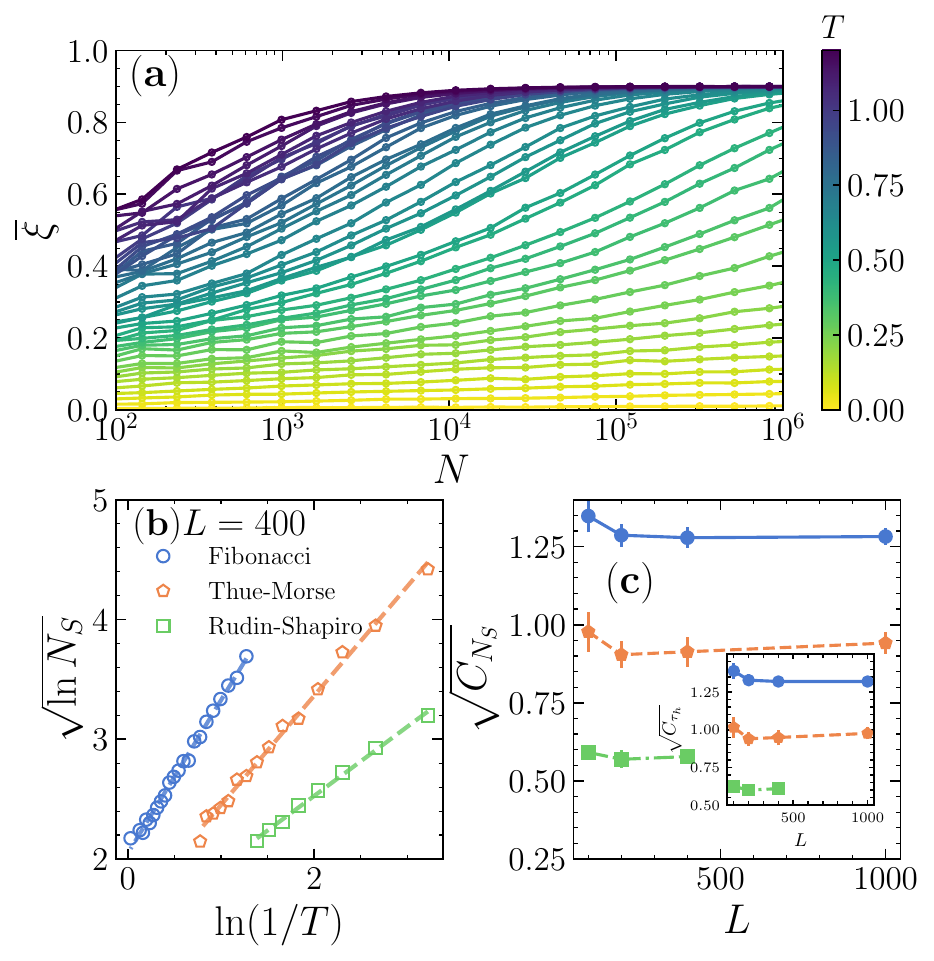}
    \caption{{\bf (a)} $\overline{\xi}$ as a function of the length of sequence $N$ at fixed disorder amplitude $\lambda = 3$, for various driving times $T$; system size is $L = 1000$. {\bf (b)} The renormalized critical length of the sequence $\sqrt{\ln N_S}$ to reach the $\overline{\xi} = 0.5$ as a function of the driving rate $\ln(1/T)$ for Fibonacci, Thue-Morse, and Rudin-Shapiro sequences (see text); here the system size is $L = 400$ and the lines are linear fittings in the range where they are displayed. {\bf (c)} The corresponding slope $\sqrt{C_{N_S}}$ obtained by linear fitting in {\bf (b)} for various sequences and system sizes $L$; the inset shows the same when recast in terms of the actual time $\tau$ (see text).}
    \label{Fig3_4}
\end{figure}

\subsection{Prethermalization}
\label{sec:prethermalization}

Previous results for the average IPR exponent in specific regimes of parameters suggest that one needs asymptotically large quasiperiodic drives to observe delocalization and, thus heating toward a featureless state. Such a \textit{preergodic} regime is not unfamiliar to isolated quantum systems driven by quasi-periodic sequences, including Fibonacci~\cite{dumitrescu2018logarithmically}, Thue-Morse~\cite{mori2021rigorous} and random-multipolar drives~\cite{zhaohongzhen2021random, zhaohongzheng2022localization}. While rigorous bounds on the heating rate have been developed for the latter two types of driving~\cite{mori2021rigorous}, it is a daunting task to obtain such bounds for generic quasi-periodic sequences.

Nonetheless, we note in what follows that the heating rate can be shown to be extremely long, even more so for Fibonacci drivings. For example, Fig.~\ref{Fig3_4}(a) shows the average IPR exponent $\overline{\xi}$ with large kick-strengths $\lambda=3$ at stroboscopic Fibonacci times (i.e., by making use of the recurrence relation $\hat{\tilde{U}}_n = \hat{\tilde{U}}_{n-1} \hat{\tilde{U}}_{n-2} $~\cite{dumitrescu2018logarithmically}, with $\hat{\tilde{U}}_{1} = \hat U_0$ and $\hat{\tilde{U}}_{2} = \hat U_0\hat U_1$) contrasting various driving periods $T$. Notably, at small values of $T$, $\overline{\xi}$ remains small even at very long sequences.

If we define a characteristic sequence length such that $\overline{\xi}=\xi_c=0.5$, we can compile what is the minimum sequence length $N_S$ that results in delocalization according to the driving time $T$ used, as shown in Fig.~\ref{Fig3_4}(b) for a lattice with $L=400$. Reference~\cite{mori2021rigorous} showed that the heating time in a Thue-Morse quasi-periodic sequence is bounded as $\tau_h \approx e^{C[\ln(1/(Tg))]^2}$, with a constant $C$ that depends on the type of driving used, and $g$ a local energy scale. Indeed, we find that our data follow such scaling form [see Fig.~\ref{Fig3_4}(b)], whose slope $\sqrt{C}$ in the delocalization time $\sqrt{\ln N_S}$ vs.~driving rate $\ln (1/T)$ plot being close to $1$ for the Thue-Morse sequence [see Fig.~\ref{Fig3_4}(c)], in agreement with the results of Refs.~\cite{mori2021rigorous, zhaohongzheng2022localization}, which employ completely different Hamiltonians than ours. 

These results, which exhibit a small system-size dependence [see again  Fig.~\ref{Fig3_4}(c)], show that the coefficient in the heating time strongly depends on the quasi-periodic sequence used, leading to Fibonacci-type drivings being the slowest in inducing delocalization. Once again, these results point out the role of the sequence's complexity in determining the fate of the localization properties: The smaller the Lempel-Ziv complexity, the longer the sequence (and consequently time) needed to induce heating. In particular, we notice that while we cast the heating times in terms of the sequence length $N_S$, we notice that they are related, i.e., 
\begin{equation}
\tau_h = \frac{N_0}{N_1+N_0}N_S T\ .
\end{equation}
For the case of the Fibonacci sequence, this can be rewritten as $\tau_h = N_S T /(\tilde\varphi+1)$, while for Thue-Morse or Rudin-Shapiro, we obtain that $\tau_h = N_ST/2$. This does not affect the scaling relation, and the slopes $\sqrt{C}$ are unchanged [see inset in Fig.~\ref{Fig3_4}(c)]. We notice that in quasi-periodic drivings in other types of Hamiltonian, exponential [$\tau_h\propto \exp(1/T)$] and stretched exponential [$\tau_h\propto \exp(\sqrt{1/T})$] scaling for the heating time have been suggested for Fibonacci and Thue-Morse sequences, respectively~\cite{das2023periodically}. We notice that while for the Fibonacci driving the $\tau_h$ scaling is sufficiently good with these other functional forms, that is not the case for the Thue-Morse and Rudin-Shapiro sequences in our settings, where the above scaling form, intermediate between algebraic and exponential fares best~\cite{mori2021rigorous}.

\subsection{Wave packet propagation}
\label{sec:wave_packet}

While the analysis of the IPR of the eigenstates of the unitary `Floquet' propagator leads to a complete characterization of the physics of the quasi-periodic kAAH, it is instructive to determine how transport occurs in practice in these conditions. Suppose one starts with a state that is initially 
localized in the middle of the lattice: $|\psi(0)\rangle = \hat{c}_{L/2}^{\dagger}|0\rangle$. After a long time, $\tau = N_0T$, the state can be expressed as $|\psi(\tau)\rangle = \hat{U}_N = \hat{U}_{b_0}\hat{U}_{b_1}\hat{U}_{b_2}\dots \hat{U}_{b_N}|\psi(0)\rangle$, and the degree of the initial wave packet spreading in position space can be quantified by the root mean square of the displacement (RMSD), defined as 
\begin{equation}
    \sigma(\tau) = \left[\sum_{n=1}^{L} (n-L/2)^2|\psi_n(\tau)|^2\right]^{1/2}.
\end{equation}
Typically, the RMSD has a power law dependence in time $\sigma(\tau) \sim \tau^{\gamma}$~\cite{qinpinquan2014dynamical, cadez2017dynamical}, wherein $\gamma = 1, 1/2,$ and $0$ indicate ballistic growth, diffusive growth, and localized behavior, respectively. Regimes with $0<\gamma<1/2$ correspond to subdiffusion, whereas  $1/2<\gamma<1$ to superdiffusion.

\begin{figure}
    \centering
    \includegraphics[width=0.99\columnwidth]{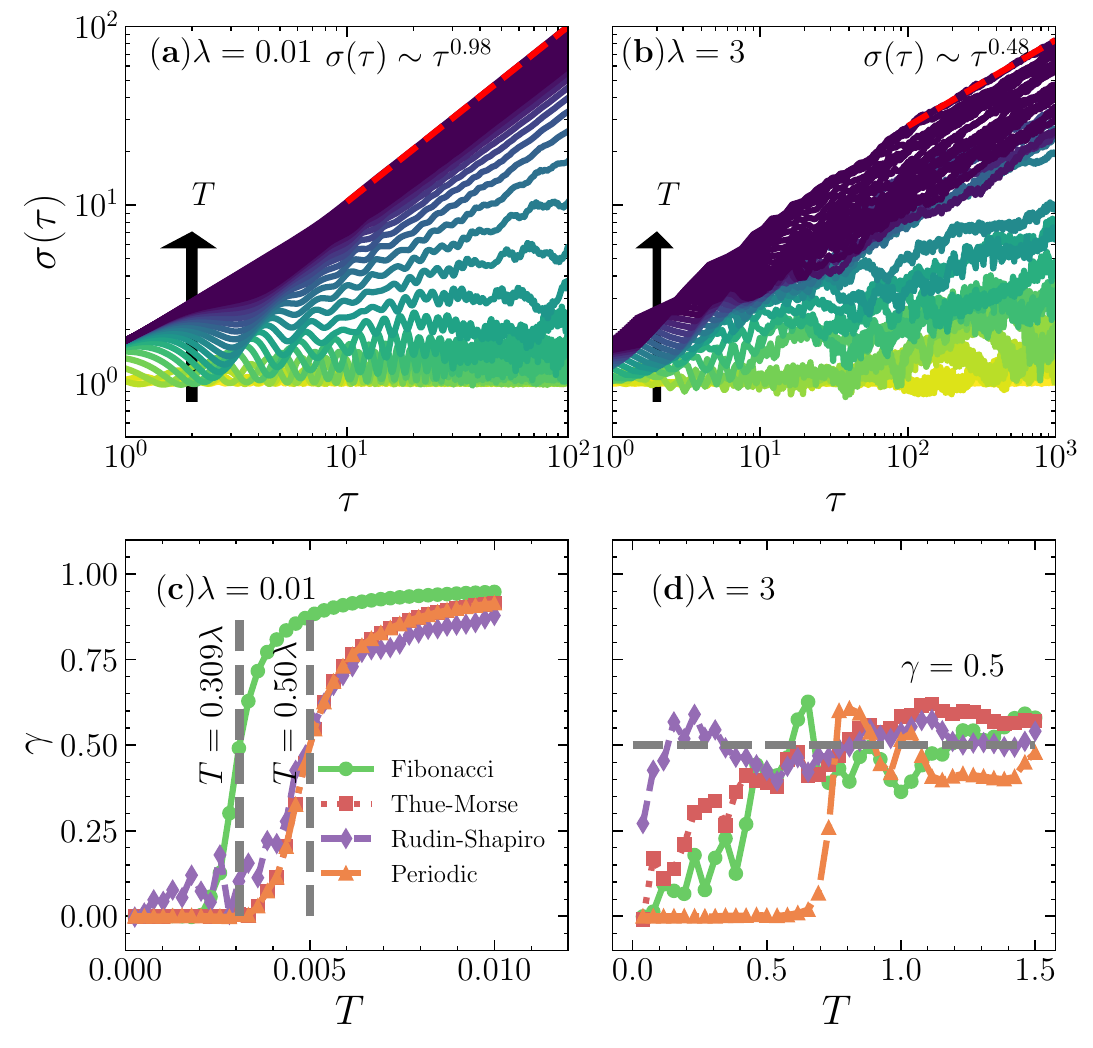}
    \caption{The RMSD $\sigma$ as a function of time $\tau$ in a Fibonacci sequence for various driving times $T$ at fixed kick amplitude (\textbf{a}) $\lambda = 0.01$ and (\textbf{b}) $\lambda = 3$. Values of $T$ in (\textbf{a}) range from $0$ to $0.01$ in steps of $2.5\times 10^{-4}$; in (\textbf{b}), from $0$ to $1.5$ in steps of $0.0375$. The red dashed line gives a power-law fit for the largest $T$ used, where we can extract the value of $\gamma$. Fitted $\gamma$ as a function of period $T$ for various sequences at fixed kick amplitude (\textbf{c}) $\lambda = 0.01$ and (\textbf{d}) $\lambda = 3$; the vertical dashed lines in the former annotate the transition point for small kick strengths, $T = N_1\lambda/(2N_0)$. The horizontal dashed line in (\textbf{d}) marks the diffusive regime where multifractality abounds.}
    \label{Fig5}
\end{figure}

In Figs.~\ref{Fig5}(a) and \ref{Fig5}(b), we present $\sigma(\tau)$ in the Fibonacci driving sequence for kick-strengths $\lambda = 0.01$ and $\lambda = 3$, respectively, with different driving times $T$. In the small kick-strength limit, the wave packet's width evolves from localization to ballistic transport with growing $T$. On the contrary, as shown in Fig.~\ref{Fig5}(b) for large kick-strengths, $\gamma$ goes from $0$ to $\simeq 0.5$, signifying a diffusive spread in this regime. Compiling these exponents  $\gamma$ for different values of $T$, we contrast in Fig.~\ref{Fig5}(c) the different sequences for $\lambda = 0.01$. In this small kick-strength case, $\gamma$ quickly evolves from 0 to 1. Still, the departure reaches $\gamma = 0.5$ in the critical point $T = N_1\lambda/(2N_0)$, for whichever driving type, quasi-periodic or periodic, where diffusive behavior characterizes the corresponding critical regime. In turn, when $\lambda = 3$, $\gamma$ for all cases changes from $0$ to $0.5$, indicating the transition between localized and multifractal states, where periodic, Fibonacci, Thue-Morse, and Rudin-Shapiro exhibit descending order of critical points with $T$ (i.e., when $\gamma$ departs from zero).

\begin{figure}
    \centering
    \includegraphics[width=0.99\columnwidth]{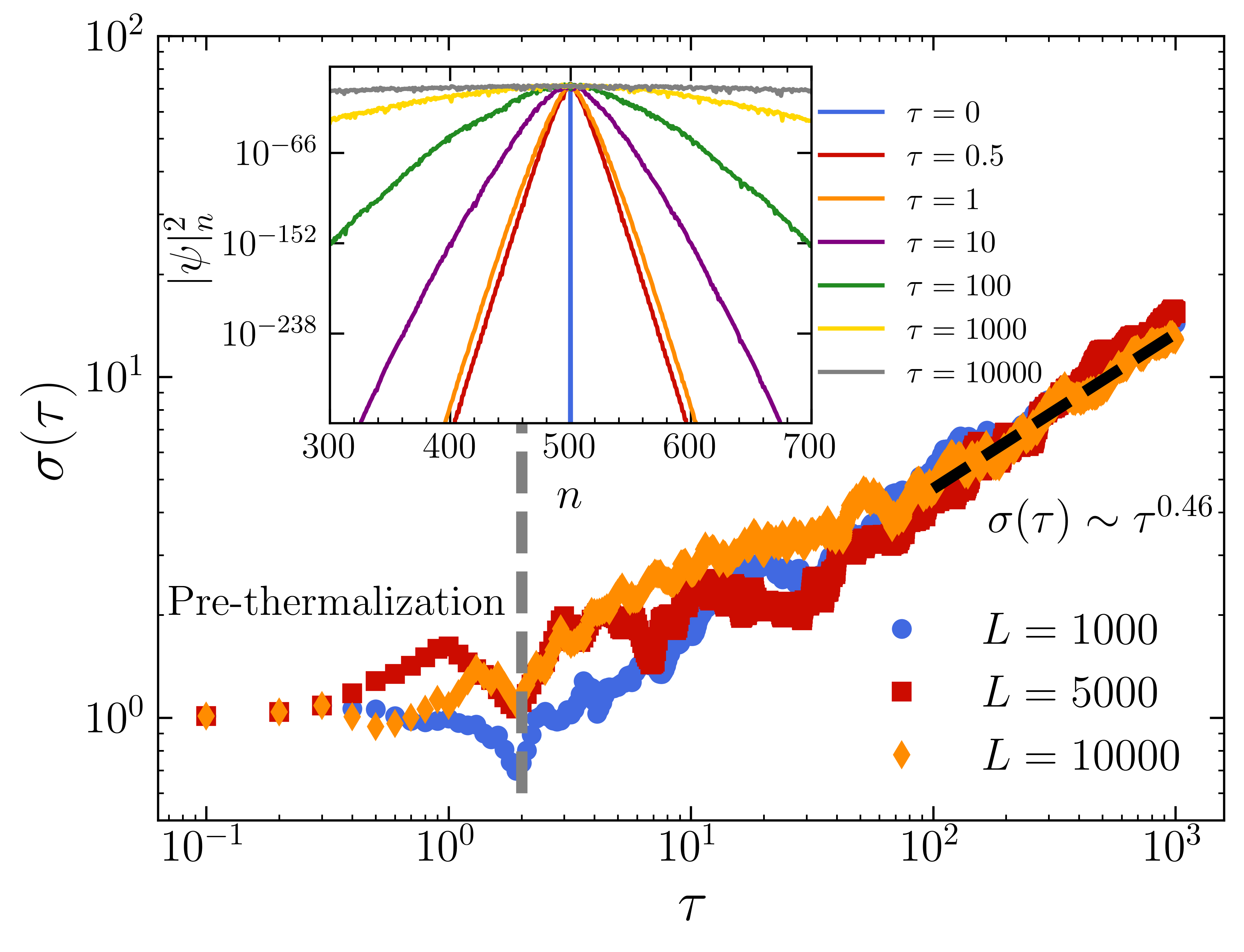}
    \caption{The RMSD $\sigma$ as a function of time $\tau$ for different system size $L$ at fixed kick amplitude $\lambda = 3$ and period $T = 0.01$. Inset shows the probability density distributions of an initially localized state after various time $\tau$.}
    \label{Fig6}
\end{figure}

A direct observation of pre-thermalization is also evident in terms of wave-packet spread. For instance, by selecting the Rudin-Shapiro sequence, the time dependence of the RMSD $\sigma(\tau)$ is shown in Fig.~\ref{Fig6}, contrasting large system sizes $L = 1000, 5000,$ and 10000 with kick amplitude $\lambda = 3$ and driving time $T = 0.1$. Before $\tau \sim 20$, the wave-packet exhibits localization and after that, it thermalizes with RMSD $\sigma(\tau) \sim \tau^{0.46}$, displaying small finite-size effects. Here, the inset shows the corresponding growth of the spatial distribution of wave functions at representative times $\tau$.

\begin{figure}
    \centering
    \includegraphics[width=0.99\columnwidth]{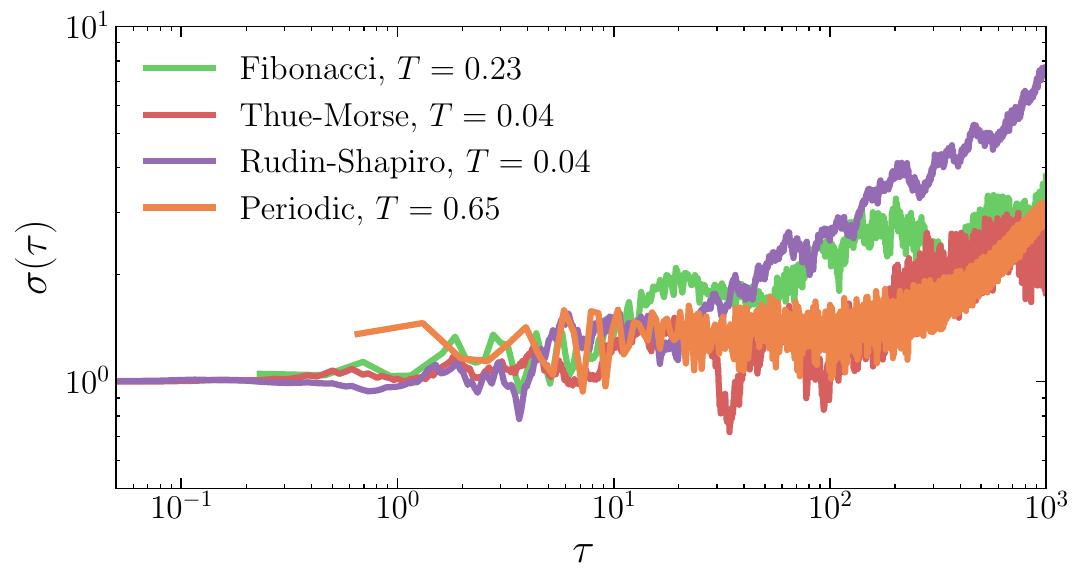}
    \caption{The $\tau$-dependence of the RMSD $\sigma$, for different quasi-periodic sequences, under strong kicks $\lambda=3$ and various driving times.}
    \label{Fig7}
\end{figure}

In fact, as our previous results have suggested, such behavior is not unique to this sequence. As shown in Fig.~\ref{Fig7}, $\sigma(\tau)$ can exhibit behavior akin to pre-termalization at large kick strengths as long as $T$ is not too large. In particular, the kAAH model can display pre-thermalization for all sequences in such regime, including periodic cases, in carefully selected sets of parameters: The less complex a sequence, the longer the driving time $T$ one can insert to observe a robust prethermalization time scale.

\section{Summary}
\label{Summary}
We have studied a family of quasi-periodic binary drivings in the kAAH model. Initially, utilizing the Floquet theorem and the BCH formula, we demonstrated the relationship between the sequence's mathematical properties, such as the fraction of its element types, $N_1/N_0$, the associated complexity, and the localization properties. In this case, we describe the critical point in the high-frequency (or short $T$) regime by a straightforward generalization of the standard dynamical localization in this model. That is, if it originally occurs at a critical point $\lambda/T=2$ for periodic drivings, it becomes $\lambda/T=2 N_0/N_1$ in quasi-periodic sequences. This is confirmed by an analysis of different quantities, such as properties of the
eigenstates of Floquet operators and the transport properties of an initially localized wave packet. Typically, the regime of validity of such critical line separating the localization and delocalization shortens when considering more complex sequences. As we analyzed in Sec.~\ref{Hamiltonian and Drive protocol}, the failure of the effective
Hamiltonian comes not only from the breakdown of the
periodicity but also from the accumulation of high-order
terms generated from the BCH formula.

We notice that both periodic or quasi-periodically kAAH models exhibit multifractal states and pre-thermalization in specific regimes of parameters. Concerning the former, we remark that the exquisite emulation of the periodic kAAH model in platforms of ultracold atoms is now a reality~\cite{shimasaki2022anomalous}, and a generalization to tackle the case of quasi-periodic sequences can be easily foreseen. Importantly, this would allow the observation of multifractality, a particular focus of that study, in a regime of parameters (smaller $T$'s and $\lambda$'s) that is more friendly to experiments using ultracold atoms owing to heating to interband transitions that naturally affect them -- see Appendix \ref{Other phase diagrams} for a detailed discussion. 

Besides that, the sequence's complexity (here quantified by the Lempel-Ziv metric) determines how fast heating occurs: the more random, the quicker the breakdown of dynamical localization when growing the sequence's length. In addition, the quasi-periodically kAAH model expands the types of driving that allow one to observe slow heating rates (prethermalization) compared to the case of completely random sequences (see Appendix \ref{Other sequences case} for an extended discussion on the regime of parameters). Lastly, we showed that the rigorously derived scaling for the heating time in a Thue-Morse sequence~\cite{mori2021rigorous} also applies to other quasi-periodic sequences. The main difference is related to the constant, seemingly a characteristic of each quasi-periodic sequence, that connects the delocalization time and the driving rate.

\begin{acknowledgments}
R.M.~acknowledges support from the NSFC Grants No.~NSAF-U2230402, No.~12111530010, No.~12222401, and No.~11974039. P.D.S. acknowledges the support from FCT through Grant No.~UID/CTM/04540/2019. Numerical simulations were performed in the Tianhe-2JK at the Beijing Computational Science Research Center.
\end{acknowledgments}

\appendix
\section{Other quasi-periodic sequences}
\label{Other sequences case}

\begin{table}
    \centering
    \setlength{\tabcolsep}{6mm}{
    \begin{tabular}{ccc}
   \toprule
   Sequence & $\mu_+$ & $N_1/N_0$\\
   \midrule
   Fibonacci & $\frac{1+\sqrt{5}}{2}$ & $\frac{1}{\mu_+} = 0.618$ \\
   Silver & $1+\sqrt{2}$& $\frac{1}{\mu_+} = 0.414$\\
   Bronze & $\frac{3+\sqrt{13}}{2}$&$\frac{1}{\mu_+} = 0.303$\\
   Copper & $2$&$\frac{1}{\mu_+} = 0.5$\\
   Nickel & $\frac{1+\sqrt{13}}{2}$&$\mu_+ - 1 = 1.303$\\
   Thue-Morse & $2$&$\frac{1}{\mu_+} = 0.5$\\
   Period-doubling & $2$&$\frac{1}{\mu_+} = 0.5$\\
   Rudin-Shapiro & $2, \sqrt{2}$&$\frac{1}{\mu_+} = \frac{1}{2} = 0.5$\\
   Paper folding & $2, 1$&$\frac{1}{\mu_+} =
   \frac{1}{2}=0.5$\\
   \bottomrule
    \end{tabular}}
    \caption{Different quasi-periodic sequences and their corresponding positive eigenvalue of substitution matrices (see Ref.~\cite{barber2008aperiodic}~, pp.~132) and the ratio of its elements $N_1/N_0$. Notice that the Rudin-Shapiro and paper folding sequences possess two positive eigenvalues, owing to their intrinsic four-element nature (see Sec.~\ref{Mathematical properties of sequences}); here, the $N_1/N_0$ is determined by the largest eigenvalue.}
    \label{tab:qp_properties}
\end{table}

\begin{figure}
    \centering
    \includegraphics[width=0.99\columnwidth]{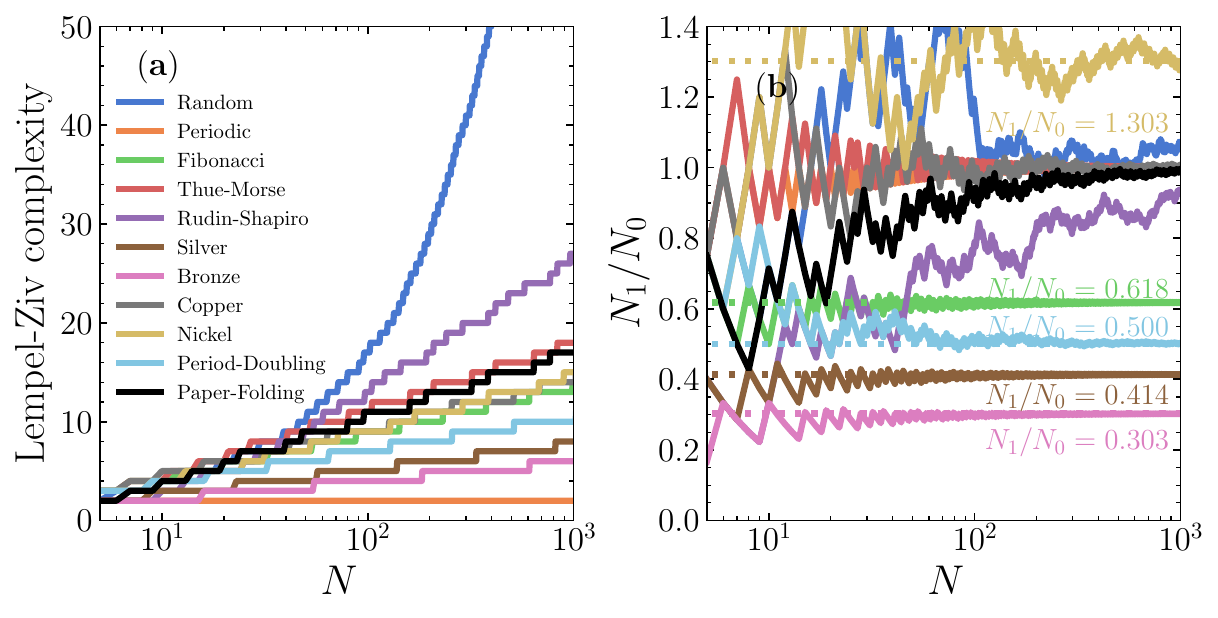}
    \caption{Extension of Fig.~\ref{Fig1} to include extra types of quasi-periodic sequences. As before, ({\bf a}) Lempel-Ziv complexity and ({\bf b}) the ratio $N_1/N_0$ of the number of its elements over a sequence length $N$. The asymptotic value of the ratio of quasi-periodic sequences can be directly extracted from the mathematical properties of the quasi-periodic sequences.}
    \label{Fig8}
\end{figure}

\begin{figure}
    \centering
    \includegraphics[width=0.99\columnwidth]{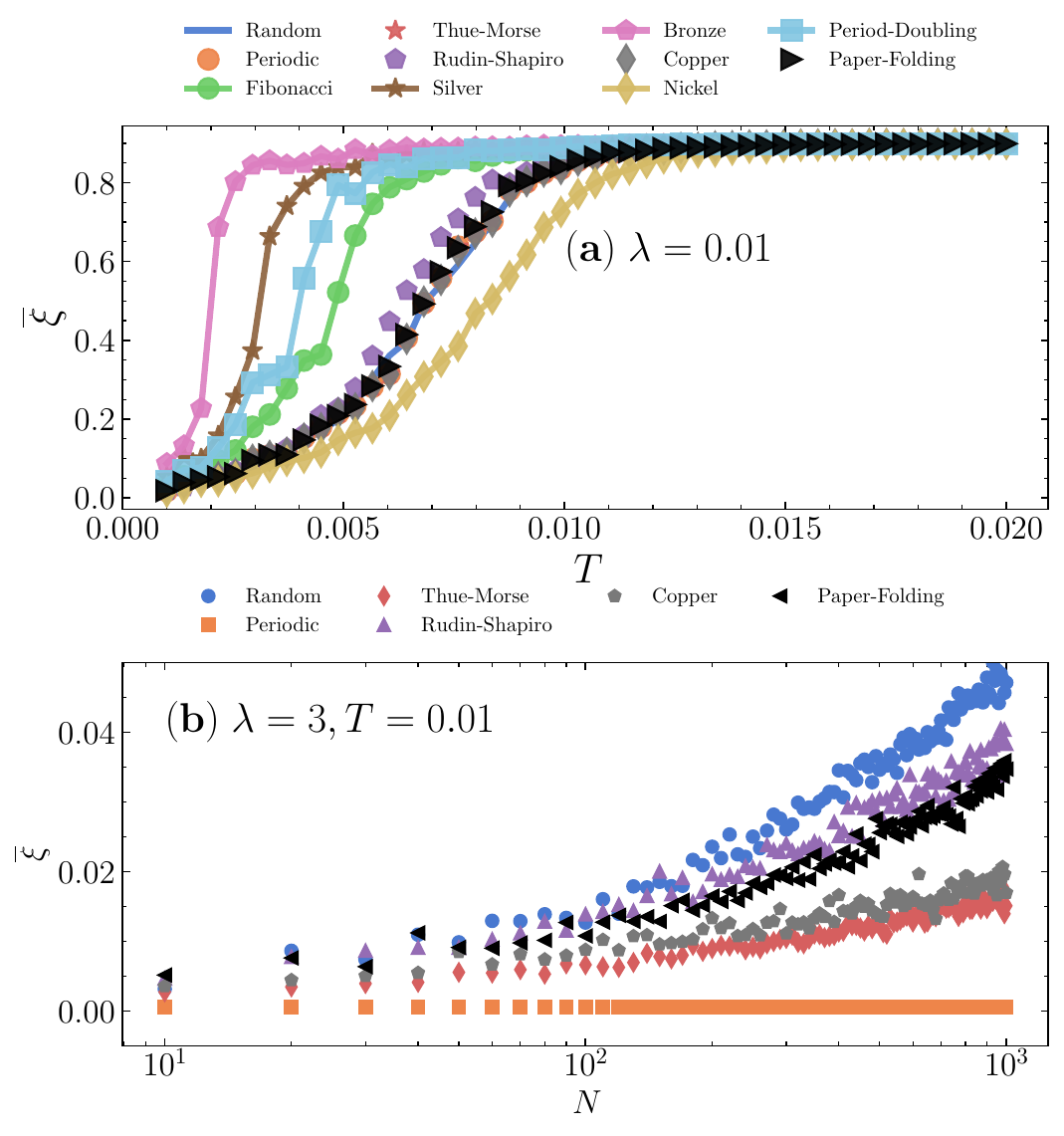}
    \caption{Similar to Fig.~\ref{Fig3} but now including extra quasi-periodic sequences as described in the legend. In ({\textbf{a}}) $\overline{\xi}$ as a function of period $T$ at fixed kick amplitude $\lambda = 0.01$; here $N = 1000$. (\textbf{b}) $\overline{\xi}$ as a function of length of sequences $N$ with kick amplitude $\lambda = 3$ and period $T = 0.01$; here we focus on sequences that exhibit $N_1/N_0=1$. Lattice size is $L= 1000$.}
    \label{Fig9}
\end{figure}

In this Appendix, we extend some of the results of the main text and check the localization properties of quasi-periodic kAAH models with various other sequences to illustrate the generality of our conclusions. All the generation rules of quasi-periodic sequences can be found in Ref.~\cite{barber2008aperiodic}. In short, one can obtain the ratio of its elements $N_1/N_0$ via the positive eigenvalues of the substitution matrices, which establish the iteration rules that govern the creation of the quasi-periodic sequences. This is summarized in Table~\ref{tab:qp_properties} and can be seen to indeed give the asymptotic numerical values of the ratio $N_1/N_0$ for large sequence sizes in Fig.~\ref{Fig8}(b). As also shown In Fig.~\ref{Fig8}(a), we also compute the corresponding Lempel-Ziv complexities: quasi-periodic sequences such as Bronze, Silver and period-doubling exhibit a smaller complexity even if compared to the Fibonacci sequence. 

According to the critical value for the extended-to-localized transition at short driving times $T$ [see Eq.~\eqref{eq:qp_critical_point}], this states that such sequences would require a much shorter driving time $T$ at the same small kick amplitude $\lambda$ to lead to delocalization. This is indeed confirmed in Fig.~\ref{Fig9}(a), in the regime where an effective Hamiltonian works well, $\lambda = 0.01$, where we show the average IPR exponent $\overline{\xi}$ vs.~$T$. That is, a small period $T$ leads to the quick build-up of $\overline{\xi}$ such that delocalization ($\overline{\xi}\to1$) ensues for the less complex sequences.

\begin{figure*}
    \centering
    \includegraphics[width=1.99\columnwidth]{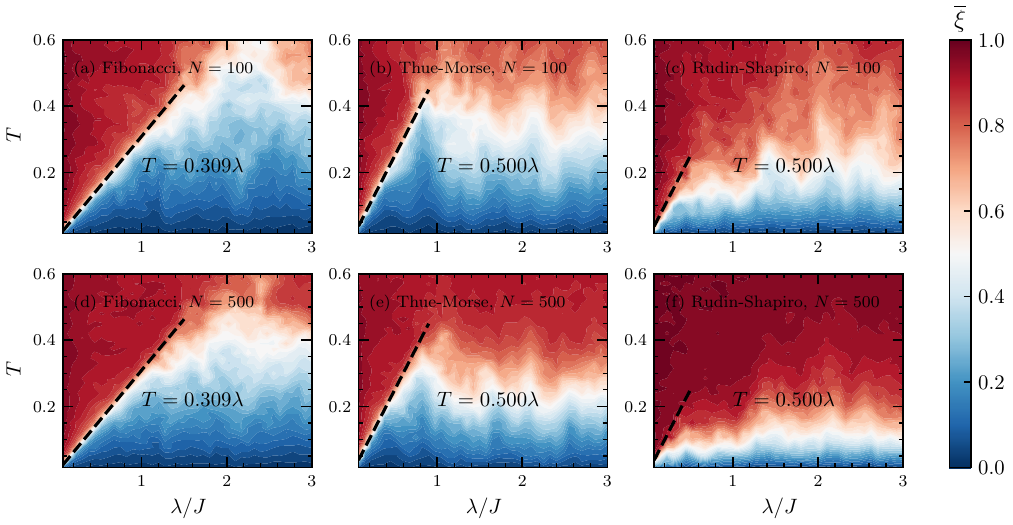}
    \caption{The average IPR exponent $\overline{\xi}$ in in the $T$--$\lambda$ plane. This phase diagram complements the one shown in Fig.~\ref{Fig2}(a) and \ref{Fig2}(b), extending it to include other sequences. In particular, [(a), (d)] show results for the Fibonacci case, [(b), (e)] for the Thue-Morse, and [(c), (f)] for the Rudin-Shapiro. Upper panels [(a)--(c)] and lower panels [(d)--(f)] give results for sequence lengths $N=100$ and $500$, respectively. Dashed lines mark the localization-delocalization transition at small $T$ and $\lambda$, given by $T = \lambda N_1/(2N_0)$. System size is $L = 100$.}
    \label{Fig7_8}
\end{figure*}

\begin{figure*}
    \centering
    \includegraphics[width = 1.99\columnwidth]{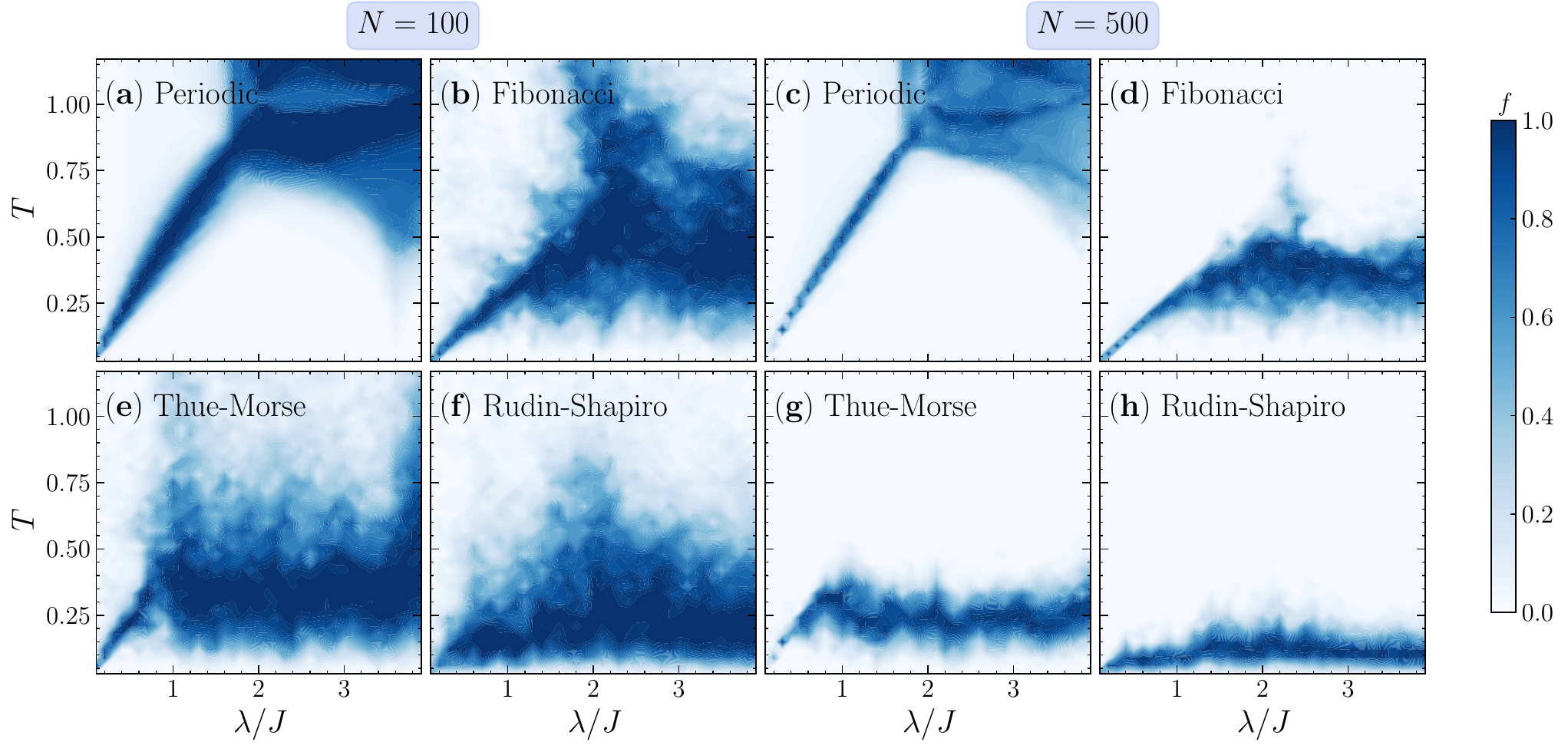}
    \caption{Fraction of the eigenstates of the Floquet operator exhibiting multifractality (see text). Two lengths of sequences are contrasted $N=100$ [(a), (b), (e), and (f)] and 500 [(c), (d), (g) and (h)]; system size is $L = 100$. Here, we further compare periodic and quasi-periodic sequences studied in the main text.}
    \label{Fig10}
\end{figure*}

In the large kick amplitude, on the other hand, as also shown in the main text, a quasi-periodic driving that is less complex typically requires a much lengthier sequence to induce delocalization, as demonstrated in Fig.~\ref{Fig9}(b); for convenience, we focus on sequences that have $N_1/N_0=1$. The observation that the rate of thermalization of localized states is intrinsically connected to the complexity can be noted by comparing paper-folding and Rudin-Shapiro, which are similarly complex, against Copper and Thue-Morse, again with close complexity: The speed of thermalization of the former two is similar and obviously greater than that of the latter two.

\section{Other phase diagrams and span of multifractality}
\label{Other phase diagrams}

In the main text we focus on the Fibonacci sequence, displaying the phase diagram given by the average IPR exponent $\overline{\xi}$ as a metric to understand (i) the localization-delocalization transition at small $T$ and $\lambda$ and (ii) to see how the delocalized region with small $\overline{\xi}$ shrinks as one progresses with longer driving sequences. Figure~\ref{Fig7_8} extends this for the Thue-Morse and Rudin-Shapiro cases. First, the line delimiting the transition, corresponding to an effective AAH time-independent Hamiltonian with potential $\frac{N_1\lambda}{N_0T}\hat H_1$ (see Eq.~\ref{eq:H_eff_qp}), is indeed generic to whichever sequence. Still, as we have predicted at the end of Sec.~\ref{Hamiltonian and Drive protocol}, such effective description narrows in the parameters space (i.e., to even smaller values of $T$ and $\lambda$) for longer driving sequences.

Besides that, the region exhibiting localization at shorter driving $T$'s, but large kick amplitudes $\lambda$ more quickly disappears for sequences with growing complexity (Fibonacci, Thue-Morse, and then Rudin-Shapiro), an indication that prethermalization is more easily seen in sequences that are less complex, as indicated by Figs.~\ref{Fig3}(b) and \ref{Fig9}(b).

One of the main goals of the recent experimental emulation of the kAAH~\cite{shimasaki2022anomalous} is the observation of multifractality in an extensive regime of parameters, as the theoretical phase diagrams suggest. Still, due to heating via interband transitions, this regime, primarily located at both large $T$ and $\lambda$, is challenging to access. In Fig.~\ref{Fig10}, we display the multifractal `phase diagram,' i.e., the fraction of eigenstates of $\hat U_N$ that exhibit IPR exponent $0.2 < \xi < 0.8$ in the $\lambda$--$T$ plane. Taking a not-so-long sequence ($N=100$), it is clear that multifractality is more easily seen at small $T$'s for quasi-periodic sequences than in the purely periodic case. Thus, this would be beneficial for experiments. Nonetheless, notice that taking much longer sequences ($N=500$), while multifractality is still seen indeed at small $T$'s, the range in the phase diagram shrinks. Consequently, balancing sequence length and complexity can substantially improve the likelihood of studying multifractal behavior in detail.

\bibliography{ref.bib}

\end{document}